\RequirePackage{ifpdf}
\documentclass[12pt,letterpaper]{article}
\usepackage{jheppub}
\usepackage{epsfig}
\usepackage[latin1]{inputenc}
\usepackage{bbm,amsfonts}
\usepackage{graphicx}
\usepackage{amssymb,amsmath}
\usepackage{fancybox}
\usepackage{verbatim}

\author[a]{Marco S. Bianchi,}
\author[b]{ Matias Leoni}
 \preprint{QMUL-PH-16-24}
\affiliation[a]{Center for Research in String Theory - School of Physics and Astronomy Queen Mary University of London, Mile End Road, London E1 4NS, UK}
\affiliation[b]{Physics Department, FCEyN-UBA \& IFIBA-CONICET\
  Ciudad Universitaria, Pabell\'on I, 1428, Buenos Aires, Argentina }

\emailAdd{m.s.bianchi@qmul.ac.uk}
\emailAdd{leoni@df.uba.ar}

\title{The canonical form has got baubles}
\abstract{The method of differential equations in canonical form has proven a powerful tool for solving multiloop Feynman integrals. In this note we test this procedure away from four dimensions. Namely, we consider the simple example of a massless doublebox, expanded in dimensional regularization around six dimensions. We achieve a canonical form for the relevant master integrals and solve them in terms of harmonic polylogarithms up to transcendental order 9. The integral basis of uniform transcendentality requires increasing indices of propagators. According to the standard graphical jargon, this amounts to decorating the integrals with baubles, like on Christmas trees, or rather loops in this case. The results can be useful for studying amplitudes in six dimensions.
\vspace{1.5cm}

\begin{center}
\includegraphics[scale=0.8]{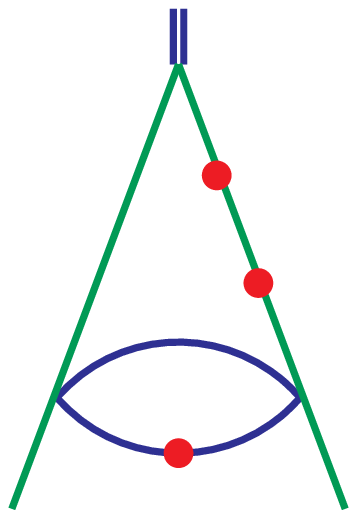}
\end{center}
}

\csname @addtoreset\endcsname{equation}{section}

\def\bseq{\begin{subequation}}  
\def\eseq{\end{subequation}}
\def\bsea{\begin{subeqnarray}}  
\def\esea{\end{subeqnarray}}

\newcommand{\beq}{\begin{equation}}
\newcommand{\bea}{\begin{eqnarray}}
\newcommand{\eea}{\end{eqnarray}}
\newcommand{\eeq}{\end{equation}}

\def\beq{\begin{equation}}
\def\eeq{\end{equation}}
\def\bea{\begin{eqnarray}}
\def\eea{\end{eqnarray}}

\begin{document}
\maketitle
\allowdisplaybreaks

\section{Introduction}

Solving Feynman integrals with differential equations is a well-established and efficient technique \cite{Kotikov:1990kg,Kotikov:1991pm,Bern:1993kr,Remiddi:1997ny,Gehrmann:1999as,Gehrmann:2000zt,Gehrmann:2001ck}.
Recently the method has been fuelled, thanks to the proposal by Henn \cite{Henn:2013pwa} of performing a particular choice of master integrals of uniform degree of transcendentality.
Such a choice casts the corresponding system of differential equations in a canonical form  where the right-hand-side is proportional to the dimensional regularization parameter. This in turn implies that the master integrals can be solved iteratively, order by order in $\epsilon$, in a quite straightforward manner (provided boundary conditions for the differential equations are specified).
The method has been applied to a plethora of contexts of phenomenological interest, that is in $d=4-2\epsilon$ dimensions \cite{Henn:2013fah,Henn:2013woa,Henn:2013nsa,Caola:2014lpa,Argeri:2014qva,Gehrmann:2014bfa,
vonManteuffel:2014mva,Dulat:2014mda,Gehrmann:2015ora,Henn:2016men,Bonciani:2016qxi,
Henn:2016kjz,Lee:2016ixa}.
On the contrary, much less applications have been performed in other dimensions (see \cite{Kozlov:2015kol,Kozlov:2016vqy}, for example).

In this short note, we take a detour from four dimensions and explore the canonical form method in higher dimensions.
In particular, we apply the method to the integrals relevant for $2\to 2$ scattering of massless particles in six dimensions up to two loops, in the planar limit.
More explicitly, we start considering the six-dimensional box as a warm-up exercise and then move to the two-loop planar doublebox.
The latter problem involves eight master integrals. We find a basis exhibiting uniform transcendentality, by showing that they obey a system of differential equations in canonical form.
The corresponding alphabet is such that the result can be expressed naturally in terms of harmonic polylogarithms HPL, whose definition we review in Appendix \ref{app:polylogs}.
We provide the necessary integration constants by requiring, order by order, that the integrals have the expected physical branch cuts. All this can be implemented in a mechanical way and we test its efficiency by pushing the solution in terms of polylogarithms up to depth 9. These results are collected in electronic format in an ancillary file \texttt{doublebox6d.m}.

We acknowledge that we carried out the relevant IBP identities reductions  \cite{Tkachov:1981wb,Chetyrkin:1981qh,Laporta:1996mq,Laporta:2001dd} using \texttt{FIRE} \cite{Smirnov:2008iw,Smirnov:2013dia,Smirnov:2014hma} and \texttt{LiteRed} \cite{Lee:2012cn,Lee:2013mka}.

\section{Warm up: the one loop case}

We first consider the one loop box integral in six space-time dimensions
\begin{equation}
G_{a_1,\dots a_4} = \frac{e^{\gamma\epsilon}}{\pi^{d/2}}\, \int d^d k_1\, \frac{1}{P_1^{a_1}\, P_2^{a_2}\, P_3^{a_3}\, P_4^{a_4}}
\end{equation}
where
\begin{equation}
P_1 = k_1^2 \quad P_2= (k_1+p_1)^2 \quad P_3= (k_1+p_1+p_2)^2 \quad
P_4= (k_1-p_4)^2
\end{equation}
We treat the integrals within dimensional regularization $d=6-2\epsilon$ and consider their Laurent expansion in the regularization parameter $\epsilon$.
Throughout the paper we work with Euclidean signature.
The kinematics of the problem allow for two independent Mandelstam invariants $s\equiv (p_1+p_2)^2$ and $t\equiv (p_2+p_3)^2$ and a single dimensionless combination thereof, which we define $x\equiv t/s$.

We derive a differential equation with respect to $x$ for this integral, which requires including two additional master integrals which are bubbles in the $s$ and $t$ channels.

In order to achieve a canonical form for the system, we choose the bubbles in such a way that they display uniform transcendentality.
By looking at their expression in terms of $\Gamma$ functions, one possible choice reads
\begin{align}
f_1 &= (1-2 \epsilon )\, \epsilon\, G_{2,0,1,0}\\
f_2 &= (1-2 \epsilon )\, \epsilon\, G_{0,2,0,1}
\end{align}
Inspecting the differential equation for the box, it turns out that the trivial corner integral suffices for having a canonical form
\begin{equation}
f_3 =  (1-2 \epsilon )\, \epsilon ^2\, (s+t)\, G_{1,1,1,1}
\end{equation}
With this choice the system of differential equations is cast in the canonical form
\begin{equation}
\partial_x \left( \begin{array}{c}
f_1\\ f_2 \\ f_3
\end{array} \right) = \epsilon\, A\, \left( \begin{array}{c}
f_1\\ f_2 \\ f_3
\end{array} \right)
\end{equation}
where
\begin{equation}
A = \left(
\begin{array}{ccc}
 0 & 0 & 0 \\
 0 & -\frac{1}{x} & 0 \\
 \frac{2}{x} & -\frac{2}{x} & \frac{1}{x+1}-\frac{1}{x} \\
\end{array}
\right)
\end{equation}
From this form it is immediate to see that the box can be expanded order by order in terms of HPL's with indices $\{0,-1\}$ of fixed transcendentality.

\section{The doublebox}

Next, we consider the class of two-loop integrals given by the following propagators
\begin{equation}
G_{a_1,\dots a_8} = \frac{e^{2\gamma\epsilon}}{(\pi^{d/2})^2}\, \int d^d k_1\, d^d k_2\, \frac{P_4^{a_4}P_6^{a_6}}{P_1^{a_1}\, P_2^{a_2}\, P_3^{a_3}\, P_5^{a_5}\, P_7^{a_7}\, P_8^{a_8}\, P_9^{a_9}}
\end{equation}
where
\begin{equation}
\begin{array}{ccccc}
P_1 = k_1^2 & \quad & P_2= (k_1+p_1)^2 & \quad & P_3= (k_1+p_1+p_2)^2 \\
P_4= (k_1-p_4)^2 & \quad & P_5 = k_2^2 & \quad & P_6= (k_2+p_1)^2\\
P_7= (k_2+p_1+p_2)^2 & \quad & P_8= (k_2-p_4)^2 & \quad & P_9=(k_1-k_2)^2
\end{array}
\end{equation}
where we assume that $a_4\geq0$ and $a_6\geq0$.
In four dimensions the problem of finding a canonical basis was solved in \cite{Henn:2013pwa}.
Here we want to achieve the same goal in $d=6-2\epsilon$.
The reduction to master integrals was solved in \cite{Smirnov:1999wz} for arbitrary dimension. In \cite{Anastasiou:2000kp} a different choice of master integrals was proposed, and the doublebox integrals where computed in six dimensions as well, thus effectively solving the problem in $6d$.
Moreover, a clever way of solving the corner doublebox in strictly six dimensions (it is both UV and IR finite and thus require no regularization) was developed in \cite{Kazakov:2014fva}, exploiting unique triangles.

\begin{figure}[h]
\includegraphics[width=\textwidth]{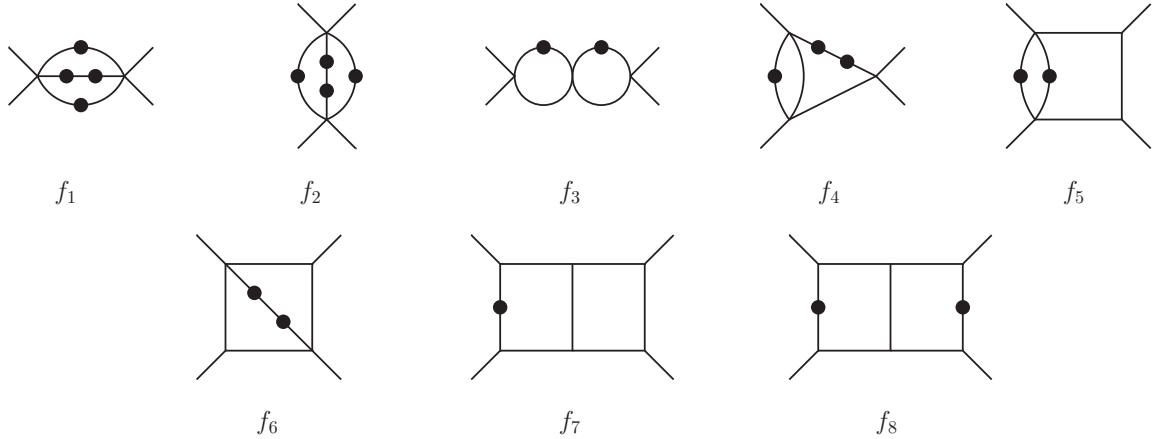}
\caption{Master integral topologies for $d=6-2\epsilon$. Dotted propagators denote raising its index by one unit. In the last two master integrals a combination of topologies is involved, which is not displayed in the cartoon.}
\label{fig:MI6}
\end{figure}

Nevertheless we revisit the problem in order to find yet another basis of master integrals, which can put the system in canonical form, with all the benefits that this carries, especially the possibility of performing higher order expansion in $\epsilon$ in a straightforward fashion.
In particular we find that the following basis does the job
\begin{align}
f_1 &= -2\, \epsilon\, (1-3 \epsilon )\, t\, G_{0,2,0,0,0,0,0,2,3} \\
f_2 &= -2\, \epsilon\, (1-3 \epsilon )\, s\, G_{0,0,2,0,2,0,0,0,3} \\
f_3 &= 4\, \epsilon ^2\, (1-2 \epsilon )^2\, G_{2,0,1,0,2,0,1,0,0}\label{master3} \\
f_4 &= -8\, \epsilon ^2\, (1-3 \epsilon )\, s\, G_{0,1,0,0,3,0,1,0,2} \\
f_5 &= -2\, \epsilon ^3\, (1-3 \epsilon ) (1-2 \epsilon )\, (s+t)\, G_{1,1,1,0,0,0,0,2,2} \\
f_6 &= \epsilon ^3\, (1-3 \epsilon )\, (s+t)\, G_{0,1,1,0,1,0,0,1,3} \\
f_7 &= \epsilon ^4\, s\, t\, (s+t)\, G_{1,2,1,0,1,0,1,2,1} - 4\, \epsilon ^2\, (1-2 \epsilon )^2\, \frac{t}{s}\, G_{2,0,1,0,2,0,1,0,0} \label{master7}\\
f_8 &= \epsilon ^4\, (2 \epsilon -1)\, s\, t\, G_{1,1,1,0,1,0,1,2,1} + \frac{1}{2}\, \epsilon ^4\, s^{2}\, t\, G_{1,2,1,0,1,0,1,2,1} - 2\, \epsilon ^3\, (1-3 \epsilon )\, s\, G_{0,1,1,0,1,0,0,1,3}
\end{align}
where the integrals are normalized with a common factor $s^{2\epsilon}$ that we did not display for conciseness.
The basis is represented graphically in Figure \ref{fig:MI6}.
A few comments are in order. The first topologies can be solved immediately in terms of $\Gamma$ functions, therefore we have just picked up a choice of indices where the expansion in $\epsilon$ displays uniform transcendentality. In particular, the factors in front of the integrals have been tuned to ensure that this is the case. Of course other choices of indices could have worked as well, ours is just intended to minimize the prefactors.
For the more complicated topologies we found no unique answer to achieve a canonical form and we present here a particular choice that we have found particularly compact, without excluding that a more convenient form could exist.
Finally we note that, as intuition might have suggested, increasing the dimension of integrations entails raising the powers of some propagators as well, in order to attain the canonical form. This behaviour can probably be studied in a more systematic manner and extended to higher loop and more complicated integrals, using maximal cuts \cite{Henn:2013pwa,Primo:2016ebd} in higher dimensions.

The matrix for the system of differential equations reads
\begin{equation}\label{eq:DE}
\partial_x\, f(\epsilon,x) = \epsilon\, A(x)\, f(\epsilon,x)
\end{equation}
\begin{equation}
A = \left(
\begin{array}{cccccccc}
 -\frac{2}{x} & 0 & 0 & 0 & 0 & 0 & 0 & 0 \\
 0 & 0 & 0 & 0 & 0 & 0 & 0 & 0 \\
 0 & 0 & 0 & 0 & 0 & 0 & 0 & 0 \\
 0 & 0 & 0 & 0 & 0 & 0 & 0 & 0 \\
 \frac{1}{4 x} & 0 & 0 & -\frac{1}{4 x} & \frac{1}{x+1}-\frac{2}{x} & 0 & 0 & 0 \\
 \frac{1}{4 x} & -\frac{1}{4 x} & 0 & 0 & 0 & \frac{2}{x+1}-\frac{2}{x} & 0 & 0 \\
 \frac{2}{x} & -\frac{1}{x} & -\frac{1}{x+1} & \frac{1}{x} & \frac{8}{x} & \frac{12}{x+1}-\frac{12}{x} & \frac{1}{x+1}-\frac{1}{x} & -\frac{2}{x} \\
 \frac{1}{2 x} & 0 & -\frac{1}{x+1} & \frac{1}{2 x} & \frac{4}{x}-\frac{2}{x+1} & 0 & \frac{1}{x+1}-\frac{1}{x} & 0 \\
\end{array}
\right)
\end{equation}
This proves that the integrals can be solved in terms of harmonic polylogarithms \cite{Remiddi:1999ew} with indices $\{0,-1\}$ at any order.

The initial condition at $\epsilon=0$ fixes the symbol of the solution completely
\begin{equation}
f^{(0)} = \{ 1,1,1,1,0,0,1,1/2 \}
\end{equation}
This can be integrated order by order in a straightforward manner.
To extract the full solution for the integrals one has to specify a boundary condition at each term in the $\epsilon$ expansion as well.
One could in principle do this by direct calculation in a special kinematic point from the Mellin-Barnes representation \cite{Smirnov:1999wz}.
Instead of doing this we follow another path, based on the properties of the Fuchsian differential equation and on the expected analytic structure of the result, along the lines of \cite{Henn:2013fah,Henn:2013nsa}.
A systematic way of performing this step is as follows.
The differential equation is given by the eight dimensional system
\begin{equation}
\frac{d}{dx}f(x,\epsilon)=\epsilon\left(\frac{M_0}{x}+\frac{M_1}{1+x}\right)\, f(x,\epsilon)
\end{equation}
We choose to integrate the system from the regular singular base point $x=0$. In order to do so we decompose
\begin{equation}
f(x,\epsilon)=\mathcal{Q}(x_,\epsilon)\, f_L(x,\epsilon)
\end{equation}
where $f_L(x,\epsilon)$ is such that
\begin{equation}
f(x,\epsilon)\to f_L(x,\epsilon),\qquad\mbox{when}\quad x\to 0
\end{equation}
This means that $f_L(x,\epsilon)$ determines the asymptotic behaviour of $f(x,\epsilon)$ at $x=0$ and it is given by
\begin{equation}\label{asymptotic}
f_L(x,\epsilon)=x^{\epsilon\,M_0}\, g(\epsilon)=\exp\left(\epsilon\, M_0\, \log x\right)\, g(\epsilon)
\end{equation}
where the vector $g(\epsilon)=(g_1(\epsilon),...,g_8(\epsilon))^t$. On the other hand $\mathcal{Q}(x,\epsilon)$ is a matrix given by
\begin{equation}
\mathcal{Q}(x,\epsilon)=\sum\limits_{n=0}^{\infty}\epsilon^n\mathcal{Q}_n(x),
\qquad\mbox{with}\quad \mathcal{Q}_0(x)=\mathcal{I},\qquad\mbox{and}\quad
\mathcal{Q}_n(0)=0\quad\mbox{for}\ n>0
\end{equation}
The terms $\mathcal{Q}_n(x)$ can be seen to satisfy the recursive differential equation
\begin{equation}\label{eq:DE0}
\frac{d}{dx}\mathcal{Q}_{n+1}(x)=\frac{[M_0,\mathcal{Q}_n(x)]}{x}
+\frac{M_1\mathcal{Q}_n(x)}{1+x}
\end{equation}
These can be easily solved in terms of harmonic polylogarithms with indices $\{0,-1\}$. The first few terms read
\begin{equation}\nonumber
\mathcal{Q}_0=\mathcal{I},\quad \mathcal{Q}_1=M_1\,H_{-1}(x),\quad
\mathcal{Q}_2=[M_0,M_1]\,H_{0,-1}(x)+M_1\, M_1\,H_{-1,-1}(x)
\end{equation}
\begin{align}
\mathcal{Q}_3=&[M_0,[M_0,M_1]]\,H_{0,0,-1}(x)+[M_0,M_1.M_1]\,H_{0,-1,-1}(x)
\nonumber\\
&+M_1\, [M_0,M_1]\,H_{-1,0,-1}(x)+M_1\, M_1\, M_1\,H_{-1,-1,-1}(x)
\end{align}
Let us expand the exponential which defines $f_L(x,\epsilon)$ in (\ref{asymptotic}). We can further decompose the result in four pieces according to the $x$-dependence:
\begin{equation}\label{asymptotic2}
f_L(x,\epsilon)={f_L}^A(\epsilon)+x^{\epsilon}{f_L}^B(\epsilon)
+x^{-2\epsilon}{f_L}^C(\epsilon)+x^{-2\epsilon}\log x\, {f_L}^D(\epsilon)
\end{equation}
where each piece ${f_L}^A$...${f_L}^D$ is a vector whose components are linear combinations of the components of $g(\epsilon)$ (for details see appendix \ref{app:asym}).
Comparing these expressions with the actual asymptotic expansion for $x\to 0$ of the first six Feynman integrals we may fix $g_1(\epsilon)$ to $g_6(\epsilon)$ exactly in terms of $\Gamma$ functions (also in appendix \ref{app:asym}). We note that this way of solving the system fixes, a priori, plenty of information about its asymptotics for $x\to 0$. In fact, for example, a careful analysis shows that thanks to the structure of the matrix $M_0$, by fixing the asymptotics of $f_2$ (which is a simple double bubble) we fix the asymptotic constant term of $f_8$ (which involves a doublebox).

Nevertheless the $x\to 0$ limit for the last two master integrals is more complicated and therefore determining $g_7(\epsilon)$ and $g_8(\epsilon)$ is not as easy as for the simpler topologies. A more efficient way of fixing these constants is by inspecting the behaviour of $f_7$ and $f_8$ at $x=-1$.
At this kinematic point we can impose the physical requirement that the integrals, being planar, do not possess branch cuts in the $u$ channel, that is for $s=-t$, $x=-1$.
Indeed the alphabet of the solution allows in principle harmonic polylogarithms with indices $-1$ in the first slots which are individually singular at that point. However, the full result for the integrals should be free of such branch points.
Since the doublebox term in $f_7$ is multiplied by $(s+t)$ and is expected to be regular in $s=-t$, we can fix $g_7(\epsilon)$, order by order, by demanding that $f_7(x)\to f_3(-1)=g_3(\epsilon)$ when $x\to -1$ ({\it c.f.} (\ref{master3}) and (\ref{master7})).
Then we can fix the $n$'th term of the expansion of $g_8(\epsilon)$ appearing at order $n$, by inspecting the analytic structure at order $n+1$ and in particular extracting the coefficients of the $\log(x+1)$ singular terms. Again, imposing that this coefficient vanishes fixes $g_8(\epsilon)$ order by order.

An even more powerful tool for fixing the constant term of $g_8(\epsilon)$ is the following. A careful analysis of the Mellin-Barnes representation of $f_7$ and $f_8$ shows, without computing the details, that in the asymptotic limit of $x\to 0$ there is no $x^\epsilon$ scaling of those two functions. On the other hand, according to (\ref{asymptotic2}), the asymptotic vector $f_L$ contains a term with the $x^{\epsilon}$ scaling that is multiplied by ${f_L}^B(\epsilon)$. Since
\begin{equation}
{f_L}^B(\epsilon)=\frac{1}{3}(
g_7(\epsilon)-2 g_8(\epsilon)-4 g_6(\epsilon)
)\
\left(
0,0,0,0,0,0,1,\!-\!1
\right)^{t},
\end{equation}
we impose that $g_8(\epsilon)=\tfrac{1}{2}g_7(\epsilon)-2 g_6(\epsilon)$ so that the $x^{\epsilon}$ scaling vanishes. Using this condition we are able to fix the integration constant of $f_8$ directly at order $n$, instead of heaving to extract the divergent behaviour at order $n+1$, as outlined above.\\

Instead of solving the system from a particular base point as an evolution operator as we did above, one could also solve \eqref{eq:DE} directly in terms of harmonic polylogarithms using their definition up to an integration constant that we can fix similarly by using the $x=-1$ limit. We compared with such an alternative method as a cross-check of our results. In performing the relevant limits we have used diffusively the \texttt{HPL.m} package \cite{Maitre:2005uu,Maitre:2007kp}.\\

Equipped with this procedure and its cross-checks we have determined a solution up to order 9 in terms of harmonic polylogarithms, which the interested reader can find in electronic format in the ancillary file \texttt{doublebox6d.m}. The results are presented in such a way to be directly fed to \texttt{HPL.m}.
As an example, here we show the result for $f_8$ at order 6
{\small \begin{align}
f^{(6)}_8 &=
-45 H_{-1,-1,-1,-1,0,0}(x)+84 H_{-1,-1,-1,0,0,0}(x)+46 H_{-1,-1,-2,0,0}(x) +\nonumber\\&
-108 H_{-1,-1,0,0,0,0}(x)
+58 H_{-1,-2,-1,0,0}(x)-88 H_{-1,-2,0,0,0}(x)-52 H_{-1,-3,0,0}(x) +\nonumber\\&+96 H_{-1,0,0,0,0,0}(x)
+44 H_{-2,-1,-1,0,0}(x)-80 H_{-2,-1,0,0,0}(x)-44 H_{-2,-2,0,0}(x) +\nonumber\\&+96 H_{-2,0,0,0,0}(x)-56 H_{-3,-1,0,0}(x)+80 H_{-3,0,0,0}(x)+48 H_{-4,0,0}(x)-80 H_{0,0,0,0,0,0}(x) +\nonumber\\&
+\pi ^2\big(-\frac{45}{2} H_{-1,-1,-1,-1}(x)+\frac{43}{2} H_{-1,-1,-1,0}(x)+23 H_{-1,-1,-2}(x)-\frac{109}{6} H_{-1,-1,0,0}(x) +\nonumber\\&+29 H_{-1,-2,-1}(x)-\frac{73}{3} H_{-1,-2,0}(x)-26 H_{-1,-3}(x)+\frac{44}{3} H_{-1,0,0,0}(x)+22 H_{-2,-1,-1}(x) +\nonumber\\&-\frac{62}{3} H_{-2,-1,0}(x)-22 H_{-2,-2}(x)+\frac{50}{3} H_{-2,0,0}(x)-28 H_{-3,-1}(x)+\frac{68}{3} H_{-3,0}(x) +\nonumber\\&+24 H_{-4}(x)-\frac{34}{3} H_{0,0,0,0}(x)\big)
+\zeta (3) \big(-45 H_{-1,-1,-1}(x)+37 H_{-1,-1,0}(x) +\nonumber\\&+58 H_{-1,-2}(x)-6 H_{-1,0,0}(x)+44 H_{-2,-1}(x)-36 H_{-2,0}(x)-56 H_{-3}(x)-\frac{20}{3} H_{0,0,0}(x)\big)
 +\nonumber\\&
+ \pi ^4 \big(-\frac{83}{120} H_{-1,-1}(x)+\frac{127}{180} H_{-1,0}(x)+\frac{59}{90} H_{-2}(x)-\frac{179}{360} H_{0,0}(x)\big)
 +\nonumber\\&
+ \zeta (5)\big(3 H_0(x)-44 H_{-1}(x))+\pi ^2 \zeta (3)\big(14  H_{-1}(x)-\frac{167}{18}  H_0(x)\big)+\frac{996240 \zeta (3)^2-107 \pi ^6}{30240}
\end{align}}
At lower orders these results can be simplified considerably and expressed in terms of ordinary polylogarithms. This might be helpful for some applications, but we have refrained from doing this in the present case, as the form in terms of HPL's makes the iterative structure of the result more manifest.
The constant for the results at order 8 features the harmonic sum $S_{5,3}(\infty)=\zeta(8)+\zeta(5,3)$, along with other products of simple zeta values. Starting from order 9 several MZV's (Multiple Zeta Values) appear, whose identities in terms of a reduced set of constants are not tabulated in \texttt{HPL.m} and that we leave indicated in the \texttt{HPL.m} format.

We have performed successful numerical checks of our results using \texttt{FIESTA} \cite{Smirnov:2008py,Smirnov:2009pb}.

\section{Conclusions}

In this note we have explored the use of differential equations in canonical form away from four dimensions, finding that the method, as expected, can be successfully applied.
In particular we have focussed on the massless planar doublebox in six dimensions. We have determined a basis of master integrals of uniform transcendentality which casts the system into canonical form and we solved the system of equations up to order 9. We have provided the necessary boundary conditions order by order, by implementing the physical requirement that the integrals be regular at $x=-1$.
As a byproduct, we have proved that these integrals can all be expressed in terms of HPL only. Apart from testing the method, the results contained in this paper might be useful for applications to scattering in six dimensions, which has been the topic of some recent papers \cite{Cheung:2009dc,Bern:2010qa,Huang:2010rn,Brandhuber:2010mm,Dennen:2010dh,Bork:2013wga,
Bork:2014nma}.
In particular, they could be used to perform tests of a fresh conjecture on a BDS exponentiation \cite{Bern:2005iz} of dual conformally invariant \cite{Drummond:2006rz,Drummond:2007au,Drummond:2007aua} amplitudes in six dimensions and a possible relation to those of ${\cal N}=4$ SYM theory \cite{Bhattacharya:2016ydi}.

Finally it would be interesting to extend the analysis of this paper to other dimensions. In particular three dimensions could be a stimulating setting, given the existence of various perturbative results on four-point loop amplitudes \cite{ABM,BLMPRS,CH,BLMPS1,BLMPS2,Bianchi:2011aa,Bianchi:2012ez,Bianchi:2013iha,Bianchi:2013pfa,Bianchi:2014iia}.

\section*{Acknowledgements}

We thank Andi Brandhuber for discussions.
MB particularly thanks Joe Hayling and Rodolfo Panerai for extra CPU's.
The work of MB was supported in part by the Science and Technology Facilities Council Consolidated Grant ST/L000415/1 \emph{String theory, gauge theory \& duality}.

\vfill
\newpage

\appendix

\section{Harmonic polylogarithms}\label{app:polylogs}

The results of the integrals of this paper are expressed in terms of harmonic polylogarithms, whose definition we review in this appendix.
Harmonic polylogarithms \cite{Remiddi:1999ew} $H_{a_1,a_2,\dots,a_n}(x)$, with indices $a_i \in \{1,0,-1\}$, are constructed in a recursive manner as follows
\begin{equation}\label{eq:HPL}
H_{a_1,a_2,\dots,a_n}(x) = \int_0^x  f_{a_1}(t)\, H_{a_2,\dots,a_n}(t)\, d t
\end{equation}
where
\begin{align}
& f_{\pm 1}(x)=\frac{1}{1 \mp x} \qquad f_0(x)=\frac{1}{x}\nonumber\\
& H_{\pm 1}(x)= \mp \log(1\mp x)\qquad H_0 (x)= \log x
\end{align}
and at least one of the indices $a_i$ is non-zero.
When all indices vanish the definition reads
\begin{equation}
H_{\underbrace{0,0,\ldots,0}_{n}}(x) = \frac{1}{n!}\log^n x
\end{equation}
We also use the so called $m$-notation \cite{Remiddi:1999ew} in which integers different form $\{-1,0,1\}$ appear as indexes of the $H$ functions. A non-null integer $m$ means that there are $|m|\!-\!1$ zeros to the left of that (non-null) index. For example $H_{-3,1,2}(x)=H_{0,0,-1,1,0,1}(x)$.

\section{System asymptotics}\label{app:asym}

The asymptotic vector is defined in (\ref{asymptotic}). We have that
\begin{equation}
f_L(x,\epsilon)={f_L}^A(\epsilon)+x^{\epsilon}{f_L}^B(\epsilon)
+x^{-2\epsilon}{f_L}^C(\epsilon)+x^{-2\epsilon}\log x\, {f_L}^D(\epsilon)
\end{equation}
with
\begin{equation}
{f_L}^A(\epsilon)=
\left(
  \begin{array}{c}
    0 \\
    g_2(\epsilon) \\
    g_3(\epsilon) \\
    g_4(\epsilon) \\
    -\tfrac{1}{8}g_4(\epsilon) \\
    -\tfrac{1}{8}g_2(\epsilon) \\
    0 \\
    \tfrac{1}{4}g_2(\epsilon) \\
  \end{array}
\right),\quad
{f_L}^B(\epsilon)=
\left(
  \begin{array}{c}
    0 \\
    0 \\
    0 \\
    0 \\
    0 \\
    0 \\
    \tfrac{1}{3}g_7(\epsilon)-\tfrac{2}{3}g_8(\epsilon)-\tfrac{4}{3}g_6(\epsilon)\\
    -\tfrac{1}{3}g_7(\epsilon)+\tfrac{2}{3}g_8(\epsilon)+\tfrac{4}{3}g_6(\epsilon)\\
  \end{array}
\right)\nonumber
\end{equation}
\begin{equation}
{f_L}^C(\epsilon)=
\left(
  \begin{array}{c}
    g_1(\epsilon) \\
    0 \\
    0 \\
    0 \\
    g_5(\epsilon) +\tfrac{1}{8}g_4(\epsilon)\\
    g_6(\epsilon) +\tfrac{1}{8}g_2(\epsilon) \\
    \tfrac{2}{3}\left(g_7(\epsilon)+g_8(\epsilon)+2g_6(\epsilon)\right) \\
    \tfrac{1}{3}\left(g_7(\epsilon)+g_8(\epsilon)-4g_6(\epsilon)
    -\tfrac{3}{4}g_2(\epsilon)\right) \\
  \end{array}
\right),\qquad\mbox{and}\nonumber
\end{equation}
\begin{equation}
{f_L}^D(\epsilon)=
\left(
  \begin{array}{c}
    0 \\
    0 \\
    0 \\
    0 \\
    \tfrac{1}{4}\epsilon g_1(\epsilon) \\
    \tfrac{1}{4}\epsilon g_1(\epsilon) \\
    \epsilon\left(2g_1(\epsilon)-g_2(\epsilon)+g_4(\epsilon)+8g_5(\epsilon)
    -8g_6(\epsilon)\right)\\
    \epsilon\left(\tfrac{1}{2}g_1(\epsilon)+\tfrac{1}{2}g_4(\epsilon)
    +4 g_5(\epsilon)-4g_6(\epsilon)
    -\tfrac{1}{2}g_2(\epsilon)\right) \\
  \end{array}
\right)
\end{equation}
Comparing these expressions with the actual asymptotic expansion for $x\to 0$ of the first six Feynman integrals we may fix $g_1(\epsilon)$-$g_6(\epsilon)$ exactly by computing them by hand. We obtain
\begin{align}
g_1(\epsilon)&=g_2(\epsilon)=e^{2\epsilon\gamma}
\frac{\Gamma^3(1-\epsilon)\Gamma(1+2\epsilon)}{\Gamma(1-3\epsilon)}\nonumber\\
g_3(\epsilon)&=\frac{e^{2 \gamma  \epsilon } \Gamma^4 (1-\epsilon ) \Gamma^2 (\epsilon +1)}{\Gamma^2 (1-2 \epsilon )}\nonumber\\
g_4(\epsilon)&=\frac{e^{2 \gamma  \epsilon }
\Gamma (1-2 \epsilon ) \Gamma (1-\epsilon )^2 \Gamma (\epsilon +1) \Gamma (2 \epsilon +1)}{\Gamma (1-3 \epsilon )}\nonumber\\
g_5(\epsilon)&=\frac{g_1(\epsilon)}{4}
\bigg(
\frac{1}{2}-\frac{\Gamma (1-2 \epsilon ) \Gamma (\epsilon +1)}{2\Gamma (1-\epsilon )}
+\nonumber\\&+\epsilon (\psi ^{(0)}(1-\epsilon )-\psi ^{(0)}(2 \epsilon +1)+\psi ^{(0)}(1-2 \epsilon )+\gamma )\bigg)\nonumber\\
g_6(\epsilon)&= \frac{g_1(\epsilon)}{4}\,\epsilon
\left(-\psi ^{(0)}(2 \epsilon +1)+2 \psi ^{(0)}(1-2 \epsilon )+\gamma \right)\nonumber
\end{align}
all six $g_i(\epsilon)$ functions have uniform transcendentality.

The functions $g_7(\epsilon)$ and $g_8(\epsilon)$ that we fixed up to order 9 using the $x\to -1$ limit read
\begin{align}
g_7(\epsilon)=&1-13 \zeta _2 \epsilon ^2-\frac{44 \zeta _3 \epsilon ^3}{3}-\frac{175 \zeta _4 \epsilon ^4}{4}+\left(\frac{362 \zeta _2 \zeta _3}{3}-\frac{292 \zeta _5}{5}\right) \epsilon ^5\nonumber\\
&+\left(\frac{977 \zeta _3^2}{9}+\frac{1423 \zeta _6}{16}\right) \epsilon ^6+\left(\frac{1598 \zeta _3 \zeta _4}{3}+\frac{3446 \zeta _2 \zeta _5}{5}-\frac{1416 \zeta _7}{7}\right) \epsilon ^7\nonumber\\
&+ \left(-48 \zeta _{5,3}-\frac{5009}{9} \zeta _2 \zeta _3^2+\frac{14318 \zeta _5 \zeta _3}{15}+\frac{479729 \zeta _8}{192}\right)\epsilon ^8+O\left(\epsilon ^9\right)
\end{align}
and
\begin{align}
g_8(\epsilon)=&\frac{1}{2}-\frac{7 \zeta _2 \epsilon ^2}{2}-\frac{16 \zeta _3 \epsilon ^3}{3}-\frac{139 \zeta _4 \epsilon ^4}{8}+\left(\frac{79 \zeta _2 \zeta _3}{3}-\frac{106 \zeta _5}{5}\right) \epsilon ^5\nonumber\\
&+\left(\frac{593 \zeta _3^2}{18}-\frac{107 \zeta _6}{32}\right) \epsilon ^6+\left(\frac{17}{90} \zeta _2 \left(402 \zeta _2 \zeta _3+918 \zeta _5\right)-\frac{484 \zeta _7}{7}\right) \epsilon ^7+ \nonumber\\& + \left( -\frac{1553 \zeta_2 \zeta_3^2}{18}+\frac{4247 \zeta_3 \zeta_5}{15}-24 \zeta_{5,3}+\frac{284969\zeta_8}{384} \right) \epsilon^8
+O\left(\epsilon ^9\right)
\end{align}
where we omit the order 9 for shortness.

\bibliographystyle{JHEP}

\bibliography{biblio}

\end{document}